\documentstyle[twoside,fleqn,espcrc2,epsf]{article}

\newcommand{\AmS}{{\protect\the\textfont2
  A\kern-.1667em\lower.5ex\hbox{M}\kern-.125emS}}

\def\lsim{\raise0.3ex\hbox{$<$\kern-0.75em\raise-1.1ex\hbox{$\sim$}}}
\def\gsim{\raise0.3ex\hbox{$>$\kern-0.75em\raise-1.1ex\hbox{$\sim$}}}

\title
{
\vspace*{-40pt}
{\normalsize \hfill UTHEP-341 } \\
\vspace*{-5pt}
{\normalsize \hfill UTCCP-P-13 } \\
\vspace*{-5pt}
{\normalsize \hfill August 1996 } \\
Scaling of the critical temperature and the quark potential\\
with a renormalization group improved SU(3) gauge action
\thanks{
{Talk presented by T. Kaneko at {\it Lattice 96}.
}}}

\author{
Y.~Iwasaki\rlap,\address{Center for Computational Physics, 
University of Tsukuba, Ibaraki 305, Japan}%
\mbox{}$^,$\address{Institute of Physics, University of Tsukuba, 
Ibaraki 305, Japan}
K.~Kanaya\rlap,$\mbox{}^{\rm a,b}$
T.~Kaneko\rlap,$\mbox{}^{\rm b}$
and
T.~Yoshi\'e$\mbox{}^{\rm a,b}$
}

\begin{document}

\begin{abstract}
We study the scaling property of 
the ratio of the critical temperature $T_c$ 
to the square root of the string tension $\sigma$ 
in the SU(3) pure gauge theory
using a renormalization group (RG) improved action.
We first determine the critical coupling $\beta_c$
on lattices with temporal extension $N_t=4$ and 6,
and then calculate the static quark potential 
at the critical couplings
on lattices at zero temperature.
The values of $T_{c}/\sqrt{\sigma}$ in the infinite volume limit
are identical 
within errors,
while they are slightly larger than 
the value extrapolated to the continuum limit 
with the standard action. 
We also note that the rotational invariance of the static quark
potential is remarkably restored in the both cases, and
that the potential $V(R)$ in physical units
scales in the whole region of $R$ investigated.
\end{abstract}

\maketitle

\section{Introduction}

In studies of lattice QCD, 
it is important to control and reduce 
finite lattice spacing effects.
Several improved actions have been proposed 
for this purpose and some of them have been tested
for the scaling behavior of the critical temperature $T_c$\cite{OIAs}.
In this work we study the scaling property of $T_{c}/\sqrt{\sigma}$
using a RG improved action proposed by one of the authors\cite{RGIA}:
\[
S_g = {1/g^2} \{c_0 \sum ({\rm plaquette}) 
               + c_1 \sum (1\times 2 {\rm \ loop})\}
\]
with $c_1=-0.331$ and $c_0=1-8c_1$. 

The outline of the calculation of 
$T_{c}/\sqrt{\sigma}$ is as follows.
First we determine the critical coupling $\beta_c$'s
on $12^3\times 4$ and $18^3\times 6$ lattices.
Then the string tensions at the two $\beta_c$'s 
are evaluated from smeared Wilson loops
on $12^3\times 24$ and $18^3\times 36$ lattices,
respectively.
The string tensions in the infinite volume limit are
estimated for the both cases,
with a finite size scaling analysis applied
using the result of $\beta_c$ on a $16^3 \times 4$
lattice.


\section{Determination of $\beta_c$}

\begin{table} [hbt]
  \caption{Parameters of the finite temperature simulations}
  \label{tab:FTPTpara}
\begin{center}
  \begin{tabular}{ccc}                   \hline

    \makebox[20mm][c]{$\beta$}        &  
    \makebox[20mm][c]{sweep}          & 
    \makebox[20mm][c]{thermalization} \\ \hline
    \multicolumn{3}{l}{$12^3\times4$} \\ 
    2.250  &  12000  &  2000          \\ 
    2.275  &  125000 & 40000          \\ 
    2.300  &  10000  &  1500          \\ \hline  
    \multicolumn{3}{l}{$16^3\times4$} \\ 
    2.283  & 220000 & 40000           \\ 
    2.290  & 240000 & 40000           \\ \hline  
    \multicolumn{3}{l}{$18^3\times6$} \\ 
    2.5000  & 120000 & 15000          \\ 
    2.5125  & 256000 & 50000          \\ 
    2.5250  & 210000 & 60000          \\ 
    2.5375  & 135000 &  5000          \\ \hline
  \end{tabular}
\end{center}
\vspace{-10mm}
\end{table}

\begin{figure}[t]
\begin{center}
\leavevmode
\epsfxsize=7.5cm 
  \epsfbox{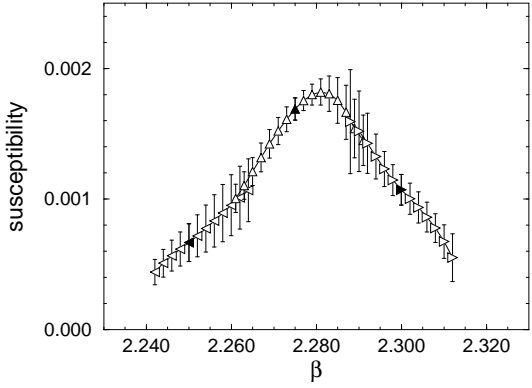}
\end{center}
\vspace{-1.5cm}
\caption{Susceptibility on the $12{^3}{\times}4$ lattice.
Filled symbols represent the location of simulation points.}
\label{fig:suscep}
\vspace{-0.4cm}
\end{figure}

The critical couplings $\beta_c$'s for the finite temperature 
phase transition in the $SU(3)$ pure gauge theory
are determined as the location of the peak 
of the susceptibility 
of the $Z(3)$ rotated Polyakov line.
Parameters of the simulations 
are summarized in Table \ref{tab:FTPTpara}.
Gauge fields are updated
by Cabibbo-Marinari-Okawa pseudo-heat bath algorithm
here and also for the calculation of the string tension.
The results of the susceptibility 
with the histogram method applied
on the $12^3\times4$ lattice 
are shown in Fig.~\ref{fig:suscep}.
The error of $\beta_c$ is determined by the jack knife 
method with bin size of 3000 sweeps
for the $12^3\times4$ and $18^3\times6$ lattices
and 6000 sweeps for the $16^3\times4$ lattice.
We have checked that the errors are stable 
for the bin size larger than these values.
We obtain 

\begin{equation}
  \beta_c = \left\{
  \begin{array}{ll}
    2.2827(16) & (12^3\times 4) \\
    2.2863(10) & (16^3\times 4) \\
    2.5157(7)  & (18^3\times 6) 
  \end{array}
\right.
\end{equation}

\section{Evaluation of the string tension}

We evaluate the string tension on a
$12^3\times 24$ ($18^3\times 36$) lattice 
at $\beta=\beta_c$ for the $12^3\times 4$ 
($18^3\times 6$) lattice.
After at least 5000 thermalization sweeps,
we measure Wilson loops at every 200 sweeps.
The spatial paths of the loops are taken to be
multiples of spatial vectors 
$(1,0,0),\ (1,1,0),\ (2,1,0),\
 (1,1,1),\ (2,1,1),\ (2,2,1) $.
The number of configurations is 200 (100) at 
$\beta_c$ for the $12^3\times 4$  
($18^3\times 6)$ lattice. 
We carry out the smearing technique proposed in Ref.\cite{StdST}
up to 30 (40) times 
on the $12^3\times 24$ ($18^3\times 36$) lattice.
The potential $V(R)$ and the overlap function $C(R)$ 
are extracted 
by fitting the Wilson loops to 
\begin{equation}
  W(R,T) = C(R) \mbox{ exp}\left[ -V(R) {\cdot} T \right].
  \label{eqn:fit1}
\end{equation}
The optimized smearing time is determined
from the condition $C(R) \simeq 1.0$.
The string tension is determined 
by fitting $V(R)$ to 
\begin{equation}
  V(R) = V_0 - \alpha/R + {\sigma}R.
  \label{eqn:fit2}
\end{equation}
We make correlated fits for the both fits.
Fitting ranges are 
\begin{equation}
  \begin{array}{lll}
    T = 3\;-\;5 & R = \sqrt{6}\;-\;4\sqrt{2} & (12{^3}{\times}24) \\
    T = 4\;-\;7 & R = 2\sqrt{3}\;-\;4\sqrt{5} & (18{^3}{\times}36) \\
  \end{array}
\end{equation}
These fitting ranges are determined by $\chi^2/df$ of the fits
and the stability of the fitted values of $\alpha$ and $\sigma$.
Note that the changes of the fitting ranges are consistent
with the change of the scale 
between $\beta=\beta_c(N_t=4)$ and $\beta_c(N_t=6)$.

\begin{figure}[t]
\begin{center}
\leavevmode
\epsfxsize=7.5cm 
  \epsfbox{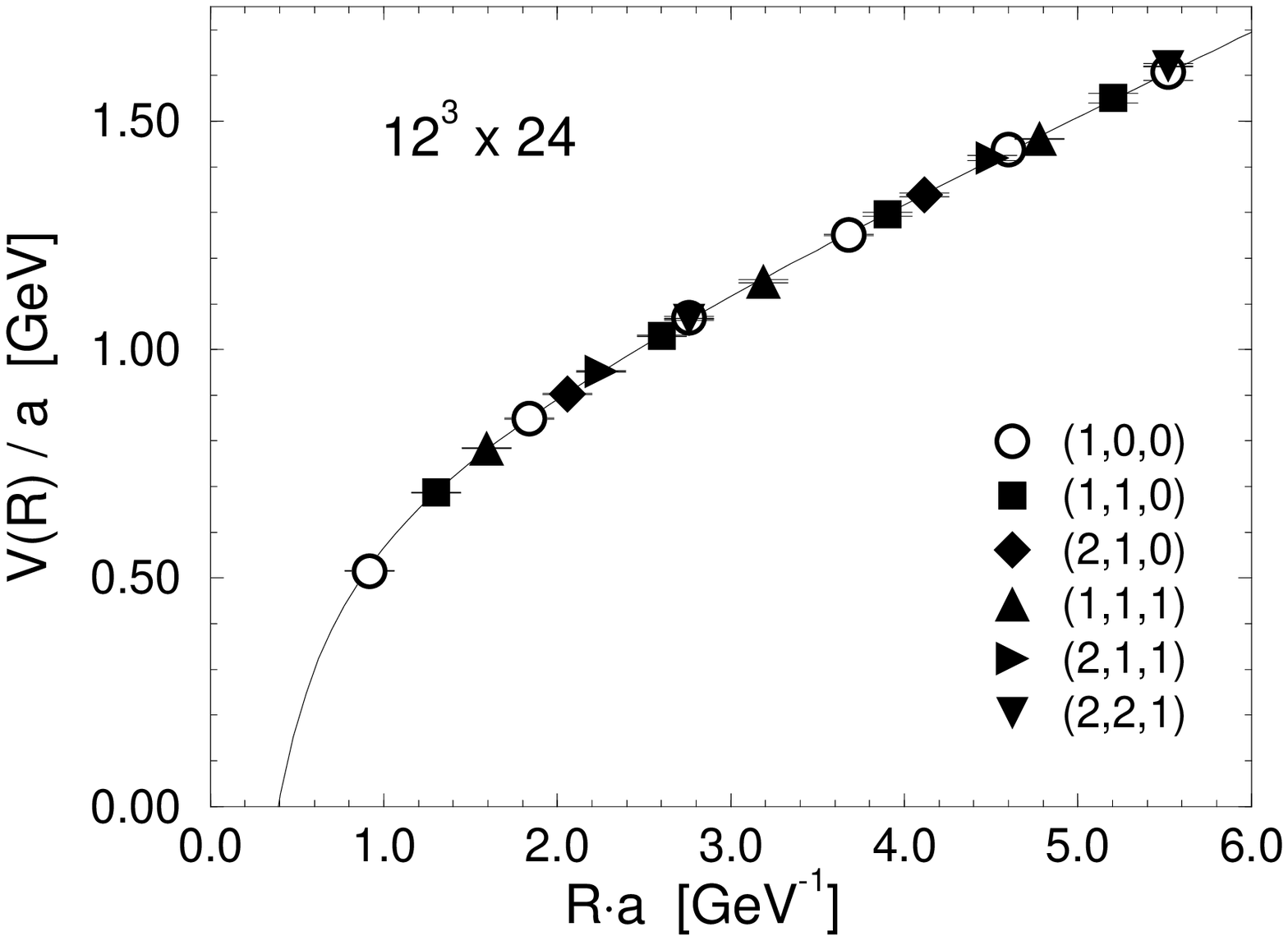}
\vspace{-0.2cm}
\epsfxsize=7.5cm 
  \epsfbox{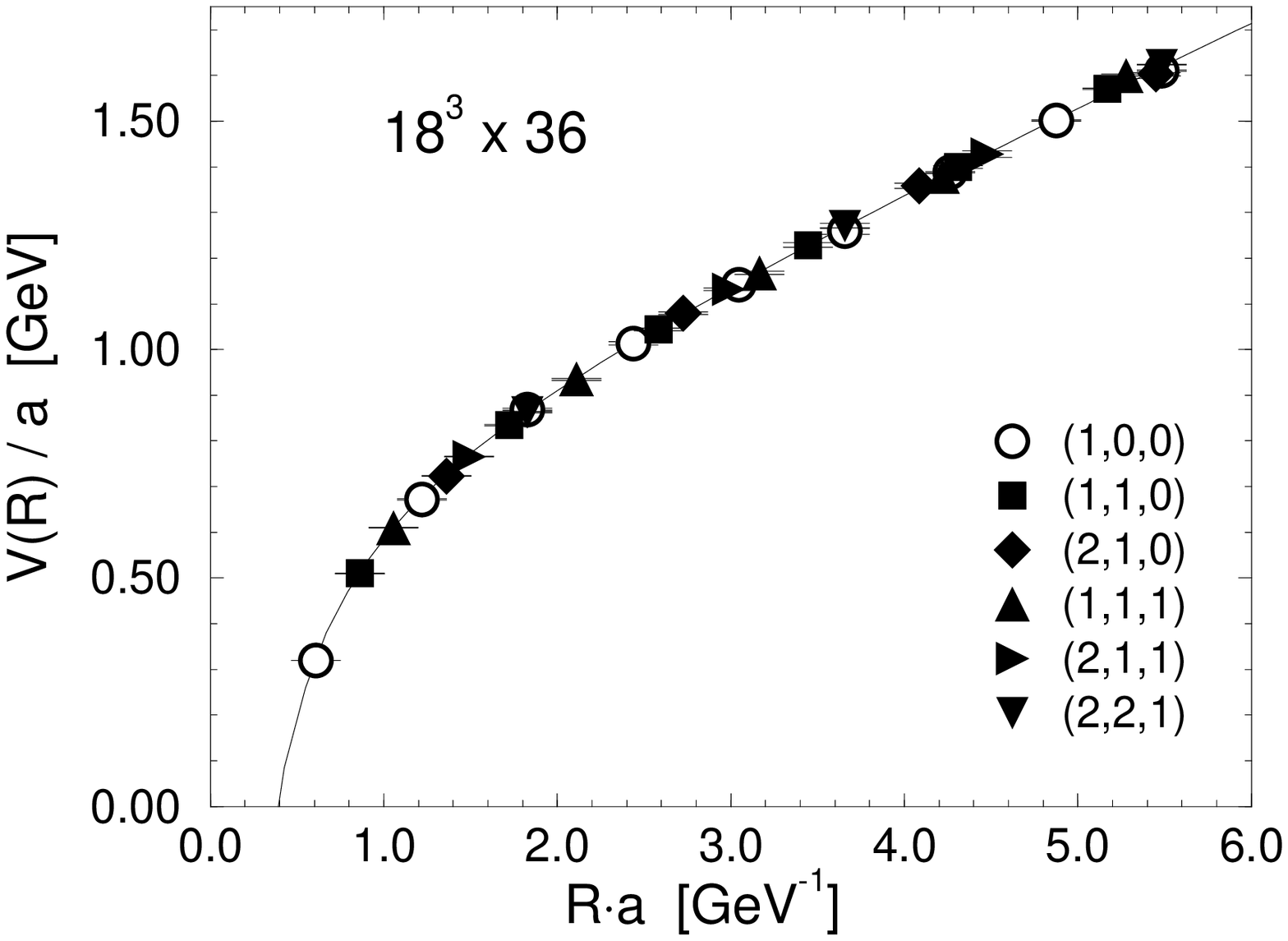}
\end{center}
\vspace{-1.5cm}
\caption{Data for the potential and its fitting curve
on the $12{^3}{\times}24$ (above) and $18{^3}{\times}36$ 
(below) lattices.}
\label{fig:VvsRds}
\vspace{-0.4cm}
\end{figure}

The values for $V(R)$ 
are plotted in Fig.~\ref{fig:VvsRds},
which are excellently fitted by the rotational invariant fitting curve, 
Eq.\ref{eqn:fit2}.
The averaged deviation is less than $0.5\%$ on the both lattices.
The results of $V_0$, $\alpha$, $\sigma$ are summarized 
in Table \ref{tbl:FitRst}.


\begin{table}[t]
  \caption{Results of $V_0$, $\alpha$, $\sigma$}
  \label{tbl:FitRst}
\vspace{-0.4cm}
\begin{center}  
  \begin{tabular}{cccc} \hline
    \makebox[07mm][c]{$N_t$}     & \makebox[15mm][c]{$V_0$}    &
    \makebox[15mm][c]{$\alpha$}  & \makebox[15mm][c]{$\sigma$} \\ 
    4  &  0.630(20)  &  0.295(14)  & 0.1493(25)  \\ 
    6  &  0.627(18)  &  0.297(19)  & 0.0655(12)  \\  \hline
  \end{tabular}
\end{center}
\vspace{-1.5cm}
\end{table}


\begin{figure}[tb]
\begin{center}
\leavevmode
\epsfxsize=7.5cm 
  \epsfbox{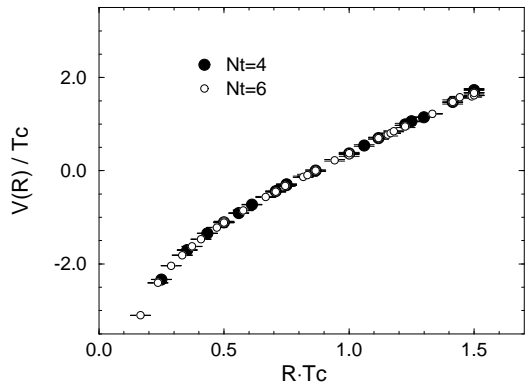}
\end{center}
\vspace{-1.5cm}  
\caption{$V/{T_c}$ vs.\ $R{\cdot}T_c$.
The constant term in the potential is fixed so that both curves
have the same value at $R{\cdot}T_c=0.866$.}
\label{fig:VvsRTc}
\vspace{-0.1cm}
\end{figure}

\section{Scaling properties}

In Fig.~\ref{fig:VvsRTc}, the values of $V(R)/T_c$ are shown
as a function of $R{\cdot}T_c$.
We note that the data obtained on the both lattices
are in excellent agreement in the whole $R{\cdot}T_c$ region.

We obtain the values of $T_c/\sqrt{\sigma}$ 
in the infinite volume limit:
\begin{equation}
T_c/\sqrt{\sigma} = \left\{
\begin{array}{ll}
0.647(+7)(5)(3) & (N_t=4) \\
0.651(+8)(6)(6) & (N_t=6)
\end{array}
\right.
\label{eqn:Tcs}
\end{equation}
The error in the first bracket is due to the shift of the critical
coupling from $\beta_c$ at $N_s/N_t=3$ to the one in the infinite volume
limit $\beta_c^{\infty}$,
the second one is the statistical error and the third one
is due to the error in the values of $\beta_c^{\infty}$. 

The error in the first bracket above has been estimated
as follows:
We have $\beta_c^\infty$ = 2.2889(29) and 2.5219(52) 
for $N_t=4$ and 6 lattices, respectively,
assuming both for $N_t=4$ and 6 lattices
a finite size scaling relation \cite{StdNt46}
obtained from
the data of $\beta_c$ on the $12^3 \times 4$ and $16^3 \times 4$
lattices:
$\beta_c(N_t,\infty) =
\beta_c(N_t,V) + {0.168(121)}\;{\cdot}\;{{N_t}^3}/{V}$.
Then the values of the string tension at $\beta_c^\infty$
are estimated assuming an exponential scaling 
$\sqrt{\sigma} = A \exp (-B \beta )$.

\begin{figure}[t]
\begin{center}
\leavevmode
\epsfxsize=7.5cm 
  \epsfbox{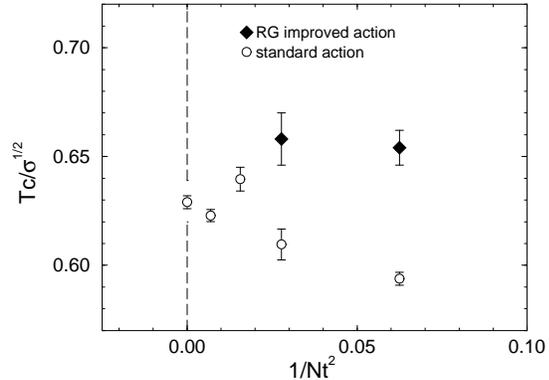}
\end{center}
\vspace{-1.5cm}
\caption{$T_c$$/$${\sigma}^{1/2}$ vs.\ $1/{{N_t}^2}$.
All symbols represent the values in the infinite volume limit.}
\label{fig:Tcs}
\vspace{-0.1cm}
\end{figure}

%
%
Our results ({\ref{eqn:Tcs}})
are shown in Fig.~\ref{fig:Tcs} together with 
the results in Ref.\cite{StdTcs2} using the standard action.
Our values are slightly larger than the value extrapolated 
to the continuum limit with the standard action $0.629(3)$.
Using the average value of (\ref{eqn:Tcs})
and the experimental value $\sigma=(420\mbox{MeV})^2$,
we estimate $T_c\sim276(3)(2)\mbox{MeV}$.
We note that $a^{-1} \approx 1.10$ GeV and 1.66 GeV, respectively,
in the $N_t=4$ and 6 cases, when we use the value of $T_c$ in
$a^{-1}=N_t T_c$.

Numerical simulations are performed with Fujitsu VPP500/30 and
HITAC H6080-FP12 at the University of Tsukuba.
This work is in part supported by 
the Grants-in-Aid of Ministry of Education,
Science and Culture (Nos.07NP0401, 07640375 and 07640376).

\end{document}